\documentstyle[12pt]{article}
\begin{document}
\begin{center}
{\large Seiberg Witten Map and the Axial Anomaly in Noncommutative
Field Theory }
\end{center}
\vskip 2cm
\begin{center}
Rabin Banerjee \footnote{e-mail:rabin@bose.res.in}\\
\vskip .5cm
S. N. Bose National Centre for Basic Sciences,\\
Block JD, Sector III, Bidhannagar, Kolkata 700098, India.\\
\vskip 1cm

 and
\vskip 1cm

Subir Ghosh \footnote{e-mail: sghosh@isical.ac.in }\\
\vskip .5cm
Physics and Applied Mathematics Unit,\\
Indian Statistical Institute, 203 B. T. Road, Kolkata 700035, India.\\
\end{center}
\vskip 1cm
{\bf Abstract :}\\
Using the point-splitting regularisation, we calculate the axial
anomaly in an arbitrary even dimensional Non-Commutative (NC)
field theory. Our result is (star) gauge invariant in its {\it
unintegrated} form, to the leading order in the NC parameter.
 Exploiting the Seiberg Witten map, this
result gets transformed to the familiar Adler-Bell-Jackiw anomaly
in ordinary space-time. Furthermore, using this map, we derive an
expression for the
 unintegrated axial anomaly for constant fields in NC space-time,
that is valid to all finite orders of the NC parameter.
\newpage
The subject of anomalies occurring in gauge theories on ordinary
space-time has a long history \cite{be}. Perhaps the most familiar
of these is the axial $U(1)$ anomaly \cite{sc}, which usually goes
by the name of Adler-Bell-Jackiw (ABJ) \cite{ad} anomaly. Ever since
the importance of Non-Commutative (NC) manifolds was realised \cite{rev}
, it has been natural to investigate the structure of anomalies in such
a setting. Various results have been reported \cite{ar} in this
context.
The structure of the axial anomaly in the NC spacetime is expected to play a
pivotal role just as the ABJ anomaly does in ordinary space-time. We feel
however that the study of this NC axial anomaly is {\it incomplete}
and a reappraisal of the problem is necessary. This is principally because
expressions for the NC anomaly are given only in their {\it integrated}
versions ($\int d^4x~<\partial _\mu \hat j^{\mu (5)}>$)
and not in their basic unintegrated forms. Strictly speaking,
therefore, the analogue of the gauge invariant ABJ anomaly in usual spacetime
with the corresponding $*$-gauge invariant anomaly in the NC spacetime is
missing.

In this paper we first show that, by properly accounting for  the
various divergences,
a $*$-gauge invariant result,
 up to the first non-trivial order in $\theta $ (the noncommutative
parameter), can be
  obtained for the $U(1)$ anomaly in its
{\it unintegrated} form. The computation is done for any even $d=2n$
dimensions. It should be stressed that the present analysis has been
done only till the first nontrivial order in $\theta$.  
For other computations in the NC spacetime
that have been done for the first nontrivial order in $\theta$, we refer
to \cite{jackiw}.

Now it is known that the Seiberg Witten (SW) map \cite{wit, wit1} connects
gauge equivalent sectors of NC theory to its conventional
counterpart. Since our result for the unintegrated anomaly is
$*$-gauge invariant, we expect that an appropriate application of
the SW map should transform this expression to the usual gauge
invariant ABJ anomaly. This is indeed so, as proved explicitly for
two and four dimensions.

Once we are convinced that it is possible to get a $*$-gauge
invariant result for the anomaly, which is connected to the usual
ABJ anomaly by the SW map, we can go further and exploit the
inverse of this map to get the unintegrated NC anomaly from the conventional
result, valid to all finite orders in $\theta $,  albeit for a
constant field tensor. This is
 because the general SW map obeys such a condition \cite{wit}.

We have also discussed another feature related to anomalies.
It is shown that a redefinition of the axial current is possible
so that it has a vanishing divergence (zero anomaly). This
modified current is, however, no longer $*$-gauge invariant. This
is the exact analogue of what happens in usual space-time.

To fix our notations and definitions, consider the ABJ anomaly
\cite{ad} in usual space-time,
\begin{equation}
<\partial _\mu j^{\mu (5)}> \equiv {\cal A} =-{{g^2}\over{16\pi
^2}}\epsilon _{\mu\nu\lambda \rho }F^{\mu\nu}F^{\lambda\rho},
\label{1}
\end{equation}
where
$j^{\mu (5)}=\bar \psi \gamma ^\mu \gamma ^5\psi $ is the
fermionic current and
$F_{\mu\nu}=\partial _\mu A_\nu-\partial _\nu A_\mu $ is
the field tensor. Obviously
the anomaly is gauge invariant under the usual Gauge Transformations
(GT) $\delta A_\mu=-\partial _\mu \Lambda $.
 
Now the corresponding expressions for the field tensor and GT in  NC
space-time are given by \cite{rev}
\footnote{We denote all variables in NC space-time by a {\it Caret}.}
$$
\hat F_{\mu\nu}=\partial _\mu \hat A_\nu-\partial _\nu \hat A_\mu
+ig[\hat A_\mu ,\hat A_\nu ]_*
$$
\begin{equation}
\delta \hat A_\mu=-\partial _\mu \hat \Lambda +ig[\hat \Lambda ,\hat A_\mu ]
_*,
\label{2}
\end{equation}
where $g$ denotes the gauge coupling and the Moyal $(*)$ bracket is defined as
$$[A,B]_*=A*B-B*A,$$
$$A(x)*B(x)=e^{{{i\theta _{\mu\nu}}\over 2}{\partial \over {\partial \eta
_\mu }}{\partial \over {\partial \zeta _\mu }}}A(x+\eta )B(x+\zeta )
\mid _{\zeta =\eta =0} =AB+{i\over 2}\theta ^{\mu\nu}\partial _\mu A
\partial _\nu B +O(\theta ^2).$$
It is easy to see that $\hat F_{\mu\nu}$ transforms covariantly under the
$*$-GT (\ref{2}),
\begin{equation}
\delta \hat F_{\mu\nu}=ig[\hat \Lambda ,  \hat F_{\mu\nu}]_*.
\label{3}
\end{equation}
The general consensus is that, adopting a $*$-gauge invariant
regularization, the $U(1)$ anomaly in NC space-time can be
computed paralleling the usual ABJ analysis. However, in order to
preserve the $*$- gauge invariance, it is essential to perform an
integration over space-time. This leads to an integrated
expression of the axial anomaly \cite{ar},
\begin{equation}
\int d^4x~ \hat{\cal A} = \int d^4x~<\partial _\mu \hat j^{\mu (5)}>=-{{g^2}\over{16\pi ^2}}
\int d^4x~ \epsilon _{\mu\nu\lambda \rho }\hat F^{\mu\nu}*\hat
F^{\lambda\rho}. \label{4}
\end{equation}
where the fermionic current is defined by a $*$-multiplication,
$\hat j^{\mu (5)}=i\bar \psi \gamma ^\mu \gamma ^5*\psi $.

Obviously, in the unintegrated version, the anomaly is no longer
$*$- gauge invariant, because of the covariant transformation law
(\ref{3}) for $\hat F^{\mu\nu}$.
To proceed with our computations of the $*$-gauge invariant anomaly,
consider the following lagrangean
\cite{ar}
\begin{equation}
{\cal L}=i\hat{\bar \psi }\gamma ^\mu *(\partial _\mu +ig\hat A_\mu )*
\hat \psi.
\label{5}
\end{equation}
The axial current in any $d = 2n$ space-time dimensions  is point-split regularized as
\begin{equation}
\hat j_\mu^{(2n+1)}(x, \epsilon )=\hat {\bar \psi }(x_+)\gamma _\mu \gamma ^{2n+1}*
\hat U(x_+,x_-)*\hat \psi (x_-),
\label{6}
\end{equation}
where $x_{\pm}=x\pm {\epsilon \over 2}$ and
the Schwinger line integral,
$$\hat U(x_+,x_-)
=(e^{-ig\int _{x_-}^{x_+}
dz^\mu \hat A_\mu(z)})_*\approx 1-ig\epsilon ^\mu \hat A_\mu +
-{{g^2}\over 2}(\epsilon .\hat A)*(\epsilon .\hat A)+O(\epsilon ^3),$$
is inserted to preserve the invariance of (\ref{6}) under the following
$*$-GT,
$$\hat \psi \rightarrow (e^{ig\hat \Lambda })_**\hat \psi ,$$
$$\hat U(x_+,x_-)\rightarrow
(e^{ig\hat \Lambda (x_+)})_**
\hat U(x_+,x_-)*
(e^{ig\hat \Lambda (x_-)})_*.$$
The equations of motion following from (\ref{5}) are
\begin{equation}
\gamma ^\mu (\partial _\mu \hat \psi +ig\hat A_\mu )*\hat \psi
=0~,~~~
 (\partial _\mu \hat {\bar \psi } -ig \hat {\bar
\psi }*\hat A_\mu )\gamma ^\mu =0. \label{6a}
\end{equation}

Taking the
divergence of the current (\ref{6}) and using the above equations
of motion,  we find that  the vacuum
expectation value, valid for the first nontrivial order in $\theta$, is given 
by,  
\begin{equation}
<\partial _\mu \hat j^{\mu (2n+1)}>=\hat{\cal A}\equiv\hat{\cal
A}_1+\hat{\cal A}_2, \label{ano}
\end{equation}
where
$$
\hat{\cal A}_1=<\hat {\bar \psi }(x_+)\gamma _{(2n+1)} \gamma _\mu
\hat \psi (x_-)>[\{ig\hat U(\hat A^\mu (x_-)-\hat A^\mu
(x_+))-\partial ^\mu \hat U\} +g\theta ^{\alpha \beta }\partial
_\beta \hat U\partial _\alpha \hat A^\mu ],$$
\begin{equation}
\hat{\cal A}_2=\theta ^{\alpha \beta }\partial _\beta [< \partial
_\alpha \hat {\bar \psi } (x_+)\gamma _{(2n+1)} \gamma _\mu \hat
\psi (x_-)>\{g(\hat A^\mu (x_+)-\hat A^\mu (x_-))-i\partial ^\mu
\hat U\}]. \label{7}
\end{equation}
In the above expressions, anti-symmetry of $\theta ^{\mu\nu}$ has
been used. It can be verified that only linear and up to quadratic
divergences survive in the VEVs in $\hat{\cal A}_1$ and $\hat{\cal
A}_2$ respectively. The presence of $\gamma _{(2n+1)}$ and the
nature of the terms besides the VEVs conspire to produce the above
result. Thus, regarding the factors multiplying the VEVs, in
$\hat{\cal A}_1$ only terms up to $O(\epsilon )$ are retained
whereas in $\hat{\cal A}_2$, terms up to $O(\epsilon ^2)$ also
contribute. This should be contrasted with the point splitting
calculation of axial anomaly in the usual case where only
$O(\epsilon )$ terms are necessary \cite{ad}. Note that in the NC anomaly
computations in the literature \cite{ar}, the $\hat{\cal
A}_2$-type of term being a surface term, is driven away by the
overall space-time integration. Herein lies the crucial difference
between our results and the usual analysis where the integrated
anomaly is computed.

We provide some details of the computation of $\hat{\cal A}_1$. The factor
in the parenthesis yields $ F^{\nu\mu }\epsilon _\nu$, while the VEV leads to 
a trace  so that,
\begin{equation}
\hat{\cal A}_1=-g\hat F^{\nu\mu }\epsilon _\nu ~tr[\gamma _
{(2n+1)}\gamma _\mu \int {{d^{2n}p}\over {(2\pi
)^{2n}}}{{e^{i\epsilon .p}\over {\gamma ^\sigma (p_\sigma -g\hat
A_\sigma )}}}]. \label{8}
\end{equation}
The propagator in NC space-time is defined as
\begin{equation}
(\gamma ^\mu (p_\mu -g\hat A_\mu ))^{-1}=
(\gamma ^\mu (p_\mu -g\hat A_\mu ))*(\gamma ^\nu (p_\nu -g\hat A_\nu ))^{-1}
*(\gamma ^\lambda (p_\lambda -g\hat A_\lambda ))^{-1}.
\label{8a}
\end{equation}
This is nothing but the conventional definition of the propagator
in ordinary space-time, with the products replaced by $*$-products
and $A_\mu \rightarrow \hat A_\mu $. Using the identity
$$
(\gamma ^\mu (p_\mu -g\hat A_\mu ))*(\gamma ^\nu (p_\nu -g\hat A_\nu ))
=(g^{\mu\nu}+{1\over 2}\sigma ^{\mu\nu })
 (p_\mu -g\hat A_\mu )*(p_\nu -g\hat A_\nu )$$
\begin{equation}
=(p_\mu -g\hat A_\mu )(p^\mu -g\hat A^\mu )+
{i\over 4}g(\sigma .\hat F) +O(\theta ^2) ,
\label{9}
\end{equation}
where $\sigma .\hat F=[\gamma _\mu ,\gamma _\nu ]\hat F^{\mu\nu}$,
and expanding the propagator in powers of $(p^2)^{-1}$, one finds
that a single term from the expansion survives, leading to,
\begin{equation}
\hat{\cal A}_1=-g\hat F^{\nu\mu }\epsilon _\nu ~tr[\gamma _
{(2n+1)}\gamma _\mu \gamma _\lambda \int {{d^{2n}p~p^\lambda
}\over {(2\pi )^{2n}p^2}} ({{-ig\sigma .\hat
F}\over{4p^2}})^{n-1}]e^{i\epsilon .p}. \label{9a}
\end{equation}
The remaining terms in the perturbative expansion of the
propagator do not contribute in $\hat{\cal A}_1$ either due to the
trace properties of $\gamma _{(2n+1)}$ or on finally taking the
$\epsilon \rightarrow 0$ limit. Using the momentum integral,
$$
\int {{d^{2n}p}\over {(p^2)^n}}e^{i\epsilon .p} =i{{\pi
^n}\over{(n-1)!}}~ln\mid \epsilon \mid ^2, $$ we arrive at
\begin{equation}
\hat{\cal A}_1={{(-i)^{n+1}g^n}\over {4^{2n-1}\pi
^n~(n-1)!}}~tr[\gamma _{2n+1}\gamma _\mu \gamma _\nu (\sigma .\hat
F)^{n-1}]\hat F^{\lambda \mu }~2{{\epsilon _\lambda \epsilon ^\nu
}\over {\mid\epsilon \mid^2}}. \label{a1}
\end{equation}
Finally, taking the trace of the $\gamma $-matrices and
incorporating the symmetric limit ${{\epsilon _\mu \epsilon _\nu
}\over {\epsilon ^2}} \mid _{\epsilon \rightarrow 0}={1\over
{2n}}\delta _{\mu\nu}$, we obtain
\begin{equation}
\hat{\cal A}_1={{(-1)^{n+1}}\over{2^{2n-1}n!}}({{g}\over \pi})^n
\epsilon _{\mu _1\nu _1...\mu _n\nu _n}\hat F_{\mu _1\nu _1}...
\hat F_{\mu _n\nu _n}. \label{10}
\end{equation}
Now we need to compute $\hat{\cal A}_2$ in (\ref{7}) where the VEV
is qualitatively different from that of $\hat{\cal A}_1$. This is
because here one of the extra derivatives that have appeared from
the $\theta $ term acts on the $\hat {\psi }$ which in the
momentum integral generates an extra momentum thus making this VEV
quadratically divergent as compared to the VEV in $\hat{\cal A}_1$
in (\ref{7}). 
The rest of the computation follows identical steps as
above and using the momentum integral,
$$\int {{d^{2n}p}\over {(p^2)^{n+1}}}e^{i\epsilon .p}
=-i{{\pi ^n}\over{4(n)!}}\mid \epsilon \mid ^2(ln\mid \epsilon
\mid ^2-1), $$ we get,
\begin{equation}
\hat{\cal A}_2=-g{{(-1)^{n+1}}\over{2^{2n-1}n!}}({{g}\over \pi})^n
\epsilon _{\mu _1\nu _1...\mu _n\nu _n}\theta ^{\alpha \beta }
\partial _\beta (\hat A_\alpha
\hat F_{\mu _1\nu _1}...
\hat F_{\mu _n\nu _n}).
\label{11}
\end{equation}
Hence the cherished expression of the NC axial anomaly is
\begin{equation}
\hat{\cal A}=\hat{\cal A}_1+\hat{\cal A}_2
=(-1)^{n+1}{1\over{2^{2n-1}n!}}({g\over \pi})^n \epsilon _{\mu
_1\nu _1...\mu _n\nu _n} [\hat F_{\mu _1\nu _1}... \hat F_{\mu
_n\nu _n}-g\theta ^{\alpha \beta }
\partial _\beta (\hat A_\alpha
\hat F_{\mu _1\nu _1}...
\hat F_{\mu _n\nu _n})].
\label{12}
\end{equation}
The second term  $\hat{\cal A}_2$, being a total derivative, is
absent in the integrated anomaly expression in \cite{ar}. Indeed,
it is precisely this piece that renders $\hat{\cal A}$ $*$-gauge
invariant, without the space-time integral prescription,
\begin{equation}
\hat \delta \hat{\cal A}=\hat \delta \hat{\cal A}_1+\hat \delta
\hat{\cal A}_2=0. \label{13}
\end{equation}

Next we relate our result for the anomaly (\ref{12}) in NC
space-time  with the usual ABJ anomaly (\ref{1})  by using the SW
map, which provides, to $O(\theta)$, the following identifications
between variables in NC and usual space-time,
$$
\hat A_\mu=A_\mu+{g\over 2}\theta ^{\alpha \beta }A_\alpha
(2\partial _\beta A_\mu -\partial _\mu A_\beta ) +O(\theta ^2),$$
\begin{equation}
\hat F_{\mu\nu}=F_{\mu\nu}-g\theta ^{\alpha \beta }(F_{\mu\alpha
}F_{\nu\beta} -A_\alpha \partial _\beta F_{\mu\nu})+O(\theta ^2).
\label{14}
\end{equation}
The SW transformation maps $*$-gauge invariant expressions (in NC
space-time) to gauge invariant expressions (in usual space-time).
Hence it is expected that our result (\ref{12}) for the NC anomaly
will reduce, through the SW map, to the anomaly in the usual
space-time. This will be explicitly shown for $1+1$ and $3+1$
dimensions.

From (\ref{12}), the two dimensional NC anomaly is,
\begin{equation}
\hat{\cal A}_2={g\over{2\pi }} \epsilon _{\mu\nu }[\hat
F^{\mu\nu}-g\theta ^{\alpha\beta }\partial _\beta
 (\hat A_\alpha \hat F^{\mu\nu }]+O(\theta ^2).
\label{15}
\end{equation}
Substituting the SW relations from (\ref{14}) we get,
$$ \hat {\cal A}\rightarrow {\cal A}
={{g}\over{2\pi }}\epsilon _{\mu\nu }F^{\mu\nu}- {{g^2}\over{2\pi
}}\epsilon _{\mu\nu } \theta _{\alpha\beta }[F^{\mu\alpha
}F^{\nu\beta}-A^\alpha
\partial ^\beta F^{\mu\nu}+\partial ^\beta (A^\alpha
F^{\mu\nu})]$$
\begin{equation}
={{g}\over{2\pi }}\epsilon _{\mu\nu }F^{\mu\nu} -{{g^2}\over{2\pi
}}\epsilon _{\mu\nu } \theta _{\alpha\beta }[F^{\mu\alpha
}F^{\nu\beta}+{1\over 2}F^{\beta\alpha} F^{\mu\nu}]. \label{an2}
\end{equation}
Exploiting the fact that in two dimensions $\theta _{\mu\nu }$ and
$\epsilon _{\mu\nu }$ are proportional, we can use the following
identity,
$$\theta_{\alpha\beta}\epsilon_{\mu\nu}\equiv
\theta\epsilon_{\alpha\beta}\epsilon_{\mu\nu}= \theta (\delta
_{\alpha\mu}\delta _{\beta\nu}-\delta_{\alpha\nu}\delta _{\beta\mu
}).$$ It is easy to see that the  $\theta $-terms drop out leaving
the usual anomaly,
\begin{equation}
{\cal A}= {{g}\over{2\pi }} \epsilon _{\mu\nu }F^{\mu\nu}.
\label{16}
\end{equation}
On the other hand, establishing the above mapping for four dimensional
space-time is more complicated. The four dimensional NC anomaly from
(\ref{12}) is
\begin{equation}
\hat {\cal A}=-{{g^2}\over{16\pi ^2}} \epsilon _{\mu \nu \rho
\lambda } [\hat F^{\mu \nu }\hat F^{\rho \lambda }-g\theta
_{\alpha \beta }
\partial ^\beta (\hat A^\alpha
\hat F^{\mu \nu }\hat F^{\rho \lambda })]. \label{17}
\end{equation}
Again substituting the NC fields in terms of the usual fields from
(\ref{14}), we find,
$$ \hat {\cal A}\rightarrow {\cal A}
=-{{g^2}\over{16\pi ^2}} \epsilon _{\mu \nu \rho \lambda }F^{\mu
\nu }F^{\rho \lambda }$$
\begin{equation}
+{{g^3}\over{16\pi ^2}} \epsilon _{\mu \nu \rho \lambda }\theta
_{\alpha\beta}(2F^{\mu\nu}F^{\rho\alpha}F^{\lambda\beta}+{1\over
2}F^{\beta\alpha}F^{\mu\nu}F^{\rho\lambda}). \label{a4}
\end{equation}
Utilising the following  tensor identity for antisymmetric
matrices $F_{\rho\lambda}$ and $\tilde F_{\rho\lambda} \equiv
2\epsilon _{\rho \lambda\mu \nu  }F^{\mu\nu} $,
$$\tilde F_{\rho\sigma}F^{\sigma\beta}F^{\rho\alpha}
=-{1\over 4}F^{\beta\alpha}\tilde F_{\mu\nu}F^{\mu\nu},$$ we find
that the $\theta $ contribution vanishes, leaving the usual ABJ
anomaly
\begin{equation}
 {\cal A}=-{{g^2}\over{16\pi ^2}} \epsilon _{\mu \nu \rho \lambda
}F^{\mu \nu }F^{\rho \lambda }. \label{aa4}
\end{equation}
The above analysis constitutes the explicit matching of the NC
anomaly and the usual anomaly via the SW map, to the leading order
in $\theta $.

Next, we move on to the construction of the modified NC current,
\begin{equation}
\hat J_\mu= \hat {\bar \psi }\gamma_\mu\gamma_{2n+1}*\hat\psi
+ \hat\Delta _\mu , \label{mod1} \end{equation}
 such that it is free of divergence anomaly,
 \begin{equation}
 \partial _\mu \hat J^\mu=0,
 \label{aam}
\end{equation}
modulo terms of $O(\theta ^2)$. Indeed, the modified current would no
longer be $*$-gauge invariant. The explicit expressions for the
modifications in $1+1$ and $3+1$ dimensions respectively  are
given below;
\begin{equation}
\hat \Delta _\mu =-{g\over \pi}\epsilon _{\mu\nu}[\hat A^\nu +{g\over 2}
\theta _{\alpha \beta }\hat A^\alpha (2\partial ^\beta \hat A^\nu -\partial
^\nu \hat A^\beta )], \label{mod}
\end{equation}
\begin{equation}
\hat \Delta _\mu ={{g^2}\over {8\pi^2}}\epsilon _{\mu\nu\rho\sigma}[\hat A^\nu
\hat F^{\rho\sigma}-g\theta _{\alpha\beta}(\hat A^\alpha
\hat F^{\beta\nu}\hat F^{\rho\sigma}-\hat A^\nu
\hat F^{\rho\alpha}\hat F^{\sigma\beta}+{1\over 2}\hat A^\alpha
\hat F^{\rho\sigma}\partial ^\nu \hat A^\beta +\hat A^\nu \hat A^\alpha\partial ^\beta
\hat F^{\rho\sigma})]. \label{mod4}
\end{equation}

Lastly, we are in a position to derive the above results, such as
the divergence anomaly, to an {\it arbitrary} finite order in the
NC parameter $\theta $. From the present analysis, 
 it can be established convincingly that, at
least to the leading order in $\theta $, even an object such as
the divergence anomaly, (which appears due to the short distance
singularity in the theory), when computed in a NC space-time,
matches identically with the result obtained from the anomaly in the
usual space-time through the SW transformation. Since the SW map,
valid for arbitrary orders in $\theta $, exists for constant field
tensor, we can get the NC divergence anomaly $\hat {\cal A}$ to
arbitrary orders in $\theta $ directly from the anomaly expression
${\cal A}$ in usual space-time, through the SW transformation
given below,
\begin{equation}
F_{\mu\nu}=\hat F_{\mu\nu}{1\over {1+g\theta .\hat F}},
 \label{arb}
\end{equation}
where $\theta .\hat F\equiv \theta ^{\alpha\beta}\hat
F_{\alpha\beta}$. Hence, for constant fields $F_{\mu\nu}$, to arbitrary
order in $\theta $, the unintegrated NC divergence anomaly in $1+1$ and $3+1$
dimensions are given by,
\begin{equation}
\hat {\cal A}={g\over {2\pi}}\epsilon_{\mu\nu}\hat
F^{\mu\nu}{1\over {1+g\theta .\hat F}},
\label{20}
\end{equation}
\begin{equation}
\hat {\cal A}=-{{g^2}\over {16\pi ^2}}\epsilon _{\mu\nu\rho\sigma}
\hat F^{\mu\nu}{1\over {1+g\theta .\hat F}}\hat
F^{\rho\sigma}{1\over {1+g\theta .\hat F}}.
\label{21}
\end{equation}

From the fact that the SW map is dimension independent, it is
obvious that all operations can be carried through in arbitrary
dimensions and subsequently the above conclusions and results will
also hold generically.

To conclude, we have found an {\it unintegrated} expression for
the axial anomaly in Non-Commutative (NC) space-time that is
$*$-gauge invariant, up to the first non-trivial order in the NC
parameter $\theta $. A point-splitting regularization has been
adopted. Contrary to the analysis in ordinary space-time,
divergences up to $O(\epsilon ^{-2})$ in the point-splitting
parameter $\epsilon $ are significant.

The structure of the NC anomaly displayed two terms; the first is
the expected modification from the usual anomaly, obtained by
replacing $F_{\mu\nu} \rightarrow \hat F_{\mu\nu} $ \cite{ar}, but
the second one is a totally new contribution. It is a total
derivative and hence vanishes if the integrated form of the
anomaly is computed \cite{ar}. 

We have also shown that an application of the Seiberg-Witten (SW)
map transforms the NC anomaly to the ordinary ABJ anomaly. This 
serves as an {\it a-posteriori} justification of our results since 
the expressions $\hat{\cal A}$ and ${\cal A}$ are
$*$ and usual gauge invariant in NC and ordinary space-time
respectively and hence should be connected through the SW map.
This is shown explicitly for $1+1$ and $3+1$ dimensions.

Having convinced ourselves that the $*$-gauge invariant anomaly in
NC space-time and the gauge invariant anomaly in ordinary
space-time are connected by the SW map, some immediate
consequences follow. The first of these is that it is possible to
redefine the axial current so that the NC anomaly vanishes.  Of
course the modified axial current is no longer $*$-gauge
invariant. Explicit expressions for such counterterms  have been
provided for $1+1$ and $3+1$ dimensions. Secondly, it was possible
to derive an unintegrated form of the NC anomaly for a constant
field, valid to all finite orders in $\theta $, by applying the
general form of the SW map in the ordinary anomaly.

Regarding future prospects, it should be possible to evaluate the
unintegrated form of the non-abelian chiral anomalies, at least to
the leading order in $\theta $, and compare the results with the
usual expressions. The implications of the SW map in this context
could also be discussed. Finally, the unintegrated divergence
anomalies can illuminate the structure of the NC commutator
anomalies along with the associated Schwinger terms. In the
ordinary case, using the point-splitting regularization, such an
approach proved useful \cite{bg}.


\newpage
\begin{thebibliography}{99}
\bibitem{be} For current reviews on this subject, see K.Fujikawa,
Int. Jour. Mod. Phys. A16 (2001) 331 and 
R.A.Bertlmann, {\it Anomalies and Quantum Field
Theory}, Oxford, U.K., Clarendon (1996).
\bibitem{sc}J.Schwinger, Phys. Rev. 82 (1951)664.
\bibitem{ad}S.L.Adler, Phys. Rev. 177 (1969) 2426; J.S.Bell And
R.Jackiw, Nuovo Cimento 60A (1969)47.
\bibitem{rev}For recent reviews, see M.R.Douglas and N.A.Nekrasov,
arXiv: HEP-TH/0108158; R.J.Szabo, arXiv: HEP-TH/0109162.
\bibitem{ar}
J.M.Gracia-Bondia and C.P.Martin, Phys. Lett. B479 (2000)321;
L.Bonora, M.Schnabl and A.Tomasiello, Phys. Lett. B485 (2000)311;
F.Ardalan and N.Sadooghi, Int. J. Mod. Phys. A16
(2001)3157,
C.P.Martin, arXiv: HEP-TH/0110046.
\bibitem{jackiw} Z. Guralnik, R. Jackiw, S.Y. Pi and A.P. Polychronakos, Phys.Lett.B517 (2001)450.
\bibitem{wit}N.Seiberg and E.Witten, JHEP 09 (1999)032.
\bibitem{wit1} For a review of the several applications of the SW map, see
C.Sochichiu, arXiv: HEP-TH/0202014.
\bibitem{bg} S.Ghosh and R.Banerjee, Z. Phys. C41 (1988)121;
Phys. Lett. B220 (1989) 581.
\end{thebibliography}
\end{document}